\documentclass[journal]{IEEEtran}
\usepackage{graphicx}
\usepackage{algorithm}
\usepackage{algorithmic}
\graphicspath{{figures/},{authors/}}
\DeclareGraphicsExtensions{.pdf,.jpg,.png}
\usepackage{caption}
\usepackage{subcaption}
\usepackage{physics}
\usepackage{float}
\usepackage{amsfonts}
\usepackage{amsmath}
\usepackage{amssymb}
\usepackage{xcolor}
\usepackage{diagbox}
\usepackage{longtable}
\usepackage{booktabs}
\usepackage{multirow}
\usepackage{cite}
\usepackage[colorlinks=true,linkcolor=blue,citecolor=blue]{hyperref} %

\usepackage{pifont}

\usepackage[capitalize]{cleveref}
\crefname{section}{Sec.}{Secs.}
\Crefname{section}{Section}{Sections}
\Crefname{table}{Table}{Tables}
\crefname{table}{Tab.}{Tabs.}

\ifCLASSINFOpdf
\else
\fi

\begin{document}
%

\title{Receive, Reason, and React: Drive as You Say with Large Language Models in Autonomous Vehicles}

\author {Can~Cui\IEEEauthorrefmark{1},~\IEEEmembership{Student Member,~IEEE,}
        Yunsheng~Ma\IEEEauthorrefmark{1},~\IEEEmembership{Student Member,~IEEE,}
        Xu~Cao,~\IEEEmembership{Member,~IEEE,}
        Wenqian~Ye,~\IEEEmembership{Member,~IEEE,}
        and~Ziran~Wang,~\IEEEmembership{Member,~IEEE}

\thanks{C. Cui, Y. Ma, and Z. Wang are with the College of Engineering, Purdue University, West Lafayette, IN 47907, USA (e-mail: cancui@purdue.edu; yunsheng@purdue.edu; ryanwang11@hotmail.com). X. Cao is with the Department of Computer Science, University of Illinois Urbana-Champaign, IL 61820, USA (e-mail: xucao2@illinois.edu). W. Ye is with the Department of Computer Science, University of Virginia, VA, 22901, USA (e-mail: wenqian@virginia.edu). X. Cao, W. Ye are also with PediaMed.AI. 

Copyright (c) 2023 IEEE. Personal use of this material is permitted. However, permission to use this material for any other purposes must be obtained from the IEEE by sending a request to pubs-permissions@ieee.org.}}

\markboth{Journal of \LaTeX\ Class Files,~Vol.~14, No.~8, August~2015}%
{Shell \MakeLowercase{\textit{et al.}}: Bare Demo of IEEEtran.cls for IEEE Journals}

\maketitle

\begingroup\renewcommand\thefootnote{\IEEEauthorrefmark{1}}
\footnotetext{These authors contributed equally.}

\begin{abstract}
The fusion of human-centric design and artificial intelligence (AI) capabilities has opened up new possibilities for next-generation autonomous vehicles that go beyond transportation. These vehicles can dynamically interact with passengers and adapt to their preferences. This paper proposes a novel framework that leverages Large Language Models (LLMs) to enhance the decision-making process in autonomous vehicles. By utilizing LLMs' linguistic and contextual understanding abilities with specialized tools, we aim to integrate the language and reasoning capabilities of LLMs into autonomous vehicles. Our research includes experiments in HighwayEnv, a collection of environments for autonomous driving and tactical decision-making tasks, to explore LLMs' interpretation, interaction, and reasoning in various scenarios. We also examine real-time personalization, demonstrating how LLMs can influence driving behaviors based on verbal commands. Our empirical results highlight the substantial advantages of utilizing chain-of-thought prompting, leading to improved driving decisions, and showing the potential for LLMs to enhance personalized driving experiences through ongoing verbal feedback. The proposed framework aims to transform autonomous vehicle operations, offering personalized support, transparent decision-making, and continuous learning to enhance safety and effectiveness. We achieve user-centric, transparent, and adaptive autonomous driving ecosystems supported by the integration of LLMs into autonomous vehicles. The experiment videos are available on {\url{https://youtube.com/playlist?list=PLgcRcf9w8BmLJi_fqTGq-7KCZsbpEIE4a&si=dhH9lgaeSmB5K94t.}}


\end{abstract}


%
\IEEEpeerreviewmaketitle




\section{Introduction}
\subsection{Background}
 Recently, Large Language Models (LLMs) have attracted significant attention. The key to their success lies in their remarkable ability to process a wide range of text inputs, including prompts, questions, dialogues, and vocabulary spanning diverse domains, resulting in significant and coherent textual outputs. LLMs store abundant information and knowledge acquired from numerous texts, much like the human brain. However, beyond their abilities in language-related tasks, LLMs have a potential that extends far beyond the realm of words and into real-world applications. Autonomous driving technologies are a very promising field and are currently drawing an increasing amount of attention \cite{10056924}. Considering the LLM's ability to emulate the human brain functions, it prompts us to ask: could we leverage the impressive capabilities of LLMs to revolutionize the future of autonomous driving?

Imagine a situation where you are in control of an autonomous vehicle, and you feel the front vehicle is too slow and desire to safely overtake that vehicle. All you have to do is speak out the command, ''Pass the vehicle in front of me." At that point, the LLMs would assess the existing conditions and safety considerations in real time, providing you with informed guidance on the feasibility and recommended actions for executing the maneuver. Furthermore, in the context of fully autonomous vehicles, the LLMs' capabilities could even extend to taking charge of the vehicle and executing the instructed commands.

The promise of integrating LLMs into autonomous driving systems is enormous. It holds the potential to enhance the safety, efficiency, and user experience of autonomous vehicles in novel ways. By gathering the vast knowledge and reasoning abilities of these models, vehicles are not only self-driving but also highly adaptive, capable of understanding and responding to the complex interactions of the road.

When integrated with LLMs, autonomous vehicles offer diverse compelling advantages over the ones that do not have LLMs enabled \cite{8944077}. These advantages extend across various aspects of functionality and performance:

\begin{itemize}
\item \textbf{Language Interaction:}  LLMs enable intuitive communication between drivers and vehicles, transforming interactions from commands to natural conversations.
\item \textbf{Contextual Understanding and Reasoning:} LLMs in vehicles offer enhanced contextual understanding from diverse sources like traffic laws and accident reports, ensuring decisions prioritize safety and regulation adherence.
\item \textbf{Zero-Shot Planning:} LLMs in vehicles can understand and reason about unfamiliar situations without prior experience, allowing vehicles to operate on uncommon scenarios confidently.
\item \textbf{Continuous Learning and Personalization:} LLMs learn and adapt continuously, providing their assistance to individual driver preferences and improving the driving experience over time.
\item \textbf{Transparency and Trust:} LLMs can articulate their decisions in simple language, enhancing trust and understanding between the technology and its users.

\end{itemize}

While LLMs have the potential to greatly enhance convenience and improve the driving experience for drivers, there are still several challenges in this technology:  LLMs lack a foundation in information from the physical world. Unlike humans, LLMs lack the ability to perceive the physical environment. In other words, these models do not possess the capability to visually perceive and interact with the world around them \cite{bender_climbing_2020}. This can render LLMs challenged in making sound decisions for the current situation, potentially leading to less-than-optimal outcomes or even hazardous consequences.

To address the challenge above, we present a novel approach where LLMs serve as the decision-making ``brain" within autonomous vehicles. Complementing this, various tools within the autonomous vehicle ecosystem, including the perception module, localization module, and in-cabin monitor, function as the vehicle's sensory ``eyes." This configuration enables LLMs to overcome the inherent limitation of not directly accessing real-time environmental information. Additionally, the vehicle's controller function as its ``hands," executing instructions derived from the LLM's decision-making process. Through receiving environmental information and drivers' commands, reasoning based on this information and human interaction, and finally making decisions, we make the autonomous driving experience that is not just technologically superior but also deeply human-centric by LLMs. This approach gains even more credibility through experiments conducted in various driving scenarios, providing proof of LLMs' revolutionalizing potential in the realm of autonomous vehicles.



\subsection{Organization}
The structure of this paper is organized as follows: In \cref{sec:review}, a comprehensive review, substantiating and validating the feasibility of employing LLMs in autonomous vehicles, is presented. Building on this foundation, we introduce the foundational concept of our human-centric LLMs framework in \cref{sec:perspective}. \cref{sec:experiments} includes experiments that verify the effectiveness and capacity of LLMs within this context. The analysis and discussion for the experiments are demonstrated in \cref{sec:discussion}. Finally, \cref{sec:conclusion} concludes the paper by summarizing our key insights and highlighting the suggested directions for future research.

\begin{figure*}[t]
    \centering
    \includegraphics[width=0.9\linewidth]{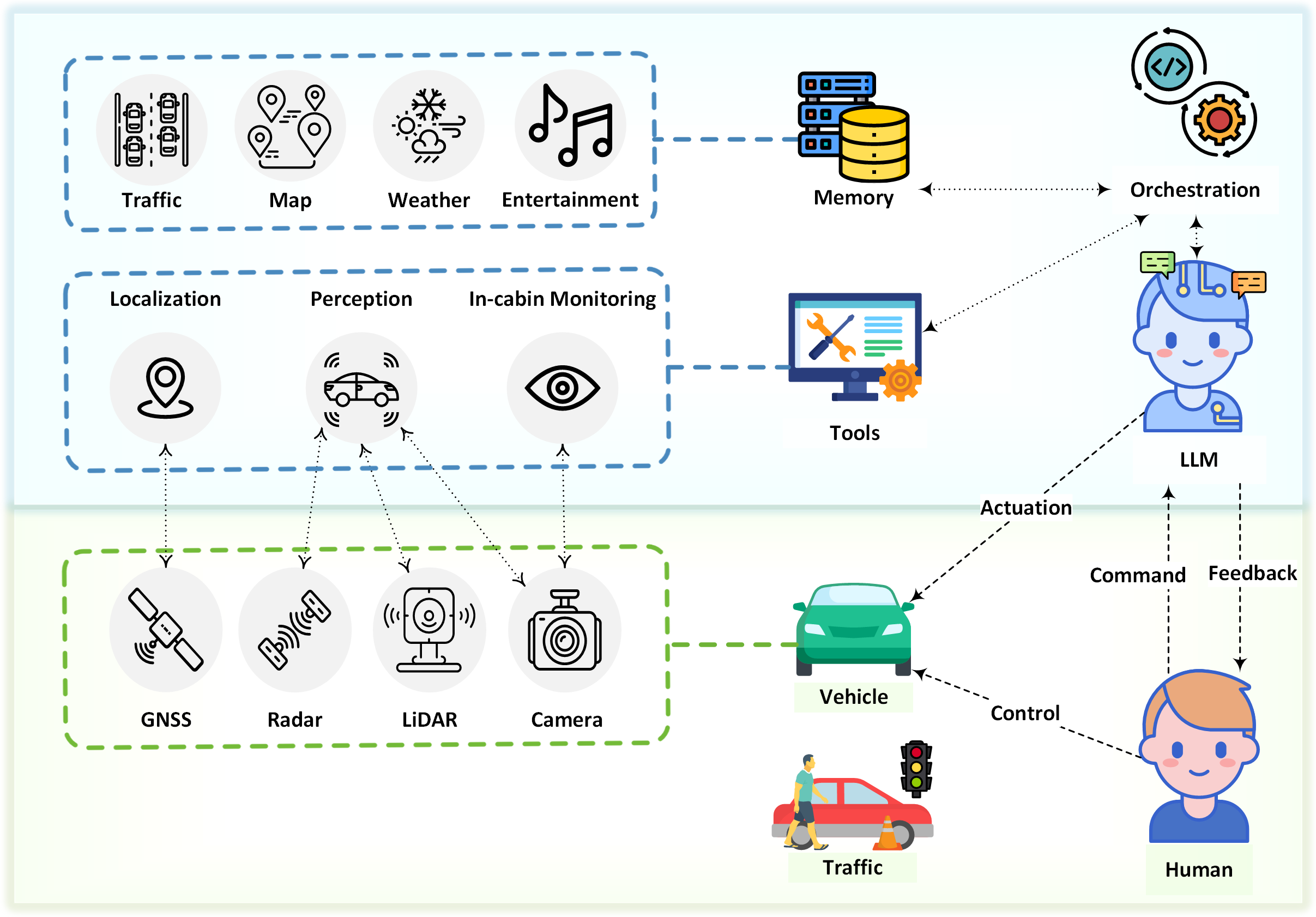}
    \caption{The human-centric LLMs-integrated framework for autonomous vehicles}
    \label{fig:framework}
\end{figure*}

\section{Related Works: Can LLMs really do this?}
\label{sec:review}
In this section, we introduce an in-depth exploration to verify the feasibility and effectiveness of integrating LLMs into human-centric autonomous driving\cite{9773320,9471770}. As LLMs offer a host of advanced linguistic capabilities, questions arise about their applicability in practical scenarios, particularly in the complex domain of autonomous vehicles. As LLMs offer a host of advanced linguistic capabilities, questions arise about their applicability in practical scenarios, particularly in the domain of autonomous driving vehicles. Through a comprehensive review of both theoretical underpinnings and real-world implementations, we try to address the fundamental question: Can LLMs really contribute to the improvement of autonomous driving by actively participating in the decision-making framework? By examining the current state of research, analyzing use cases, and considering potential challenges, this section aims to provide a thorough assessment of the extent to which LLMs can bring to the landscape of human-centric autonomous driving.

\subsection{Adaptive Techniques and Human-Centric Refinements for LLMs}
Parameter-efficient fine tuning (PEFT) is a crucial technique used to adapt pre-trained language models (LLMs) to specialized downstream applications~\cite{chung_scaling_2022,fu_effectiveness_2023,hu_lora_2022,dettmers_qlora_2023,lester_power_2021}. Hu et al.~\cite{hu_lora_2022} proposed utilizing low-rank decomposition matrices to reduce the number of trainable parameters needed for fine-tuning language models. Lester et al.~\cite{lester_power_2021} explore prompt tuning, a method for conditioning language models with learned soft prompts, which achieves competitive performance compared to full fine-tuning and enables model reuse for many tasks. These PEFT techniques offer valuable tools for adapting LLMs to autonomous driving tasks.

Reinforcement Learning from Human Feedback (RLHF) \cite{stiennon_learning_2020,ouyang_training_2022,schulman_proximal_2017,bai_constitutional_2022,rafailov_direct_2023} has emerged as a key strategy for fine-tuning LLM systems to align more closely with human preferences. Ouyang et al.~\cite{ouyang_training_2022} introduce a human-in-the-loop process to create a model that better follows instructions. Bai et al.~\cite{bai_constitutional_2022} propose a method for training a harmless AI assistant without human labels, providing better control over AI behavior with minimal human input. These approaches hold significant promise for developing LLMs for autonomous driving applications, as they can contribute in two dimensions. Firstly, they can ensure that LLMs avoid making decisions that may be illegal or unethical. Secondly, these methodologies enable LLMs to continually adapt and align their decision-making processes with user preferences, enhancing personalization and trust in autonomous vehicles.

LLM-based autonomous driving applications can also benefit from advanced prompting techniques~\cite{cui_drive_2023,kojima_large_2022,yao_react_2023,wang_self-consistency_2023,besta_graph_2023}. A good example is Chain-of-thought prompting~\cite{wei_chain--thought_2022}, whichimproves LLMs' ability to perform complex reasoning. Gao et al. \cite{gao_pal_2023} propose an approach that uses LLMs to read natural language problems and generate programs as intermediate reasoning steps. Yao et al. \cite{yao_react_2023} present a new prompting technique that allows LLMs to make decisions about how to interact with external APIs. These methods provide a solid foundation and potential direction for the development of LLMs for autonomous driving applications with two significant benefits: (1) Using these tools to enhance LLMs' reasoning capabilities, particularly in complex, multi-step scenarios. (2) Using these techniques to improve the adaptability and versatility of LLMs, key attributes for autonomous driving systems interfacing with various tools and data sources.

\subsection{Advancements in LLMs: Implications for Autonomous Driving Decision-Making}
Recent Research has shown that LLMs can perform well in most commonsense tasks \cite{bian_chatgpt_2023-1}, which means it has the potential to make wise and feasible decisions in autonomous driving scenarios. The utilization of LLMs in the context of autonomous driving presents a captivating and potentially transformative direction for research. Recent investigations have brought light on the diverse ways in which LLMs can profoundly impact the landscape of autonomous vehicles. For instance, the study conducted by \cite{nay_law_2022} highlights the promise of AI-infused with legal knowledge, offering the potential to avert legal transgressions in autonomous driving scenarios, thereby contributing to the establishment of a safer AI-driven environment. Additionally, \cite{zheng_trafficsafetygpt_2023} demonstrates that LLMs possess the capability to learn from local laws, and accident reports, and effectively contribute to reducing accident rates, thus enhancing the safety of autonomous driving.

The application of LLMs to decision-making in autonomous driving is notably explored by PaLM \cite{chowdhery_palm_2022}. They demonstrate that LLMs exhibit a capacity to effectively tackle complex reasoning tasks and exceed the performance of humans. Such a finding carries significant implications, hinting at LLMs' remarkable ability to navigate complex scenarios, make astute judgments, and potentially lay the groundwork for optimal decision-making in autonomous vehicles.


The adaptive capabilities of LLMs are showcased in various ways. \cite{kojima_large_2023} underscores LLMs' proficiency in zero-shot reasoning, enabling them to deal with novel and unfamiliar situations, a vital feature for autonomous vehicles operating in dynamic environments. The study by \cite{chung_scaling_2022} exemplifies that LLMs can be fine-tuned to exhibit enhanced performance, particularly in tasks with limited training data.

Additionally, LLMs have shown great potential in both transportation and robotics areas, as highlighted by \cite{zheng_chatgpt_2023}, and \cite{vemprala_chatgpt_2023} respectively. They reveal LLMs' prowess in tasks such as zero-shot planning and interactive conversations, even facilitating interaction with perception-action-based API libraries, an attribute that aligns with the demands of autonomous vehicles.

Furthermore, the work \cite{wang_voyager_2023} demonstrates LLMs' potential for continuous learning, which is of paramount importance for adapting to evolving road conditions and enhancing performance over time.

Incorporating insights from studies such as \cite{driess_palm-e_2023} is essential. Their investigation introduces embodied language models capable of assimilating real-world sensor data, thus bridging the gap between perception and language. This development lays the foundation for potential advancements in autonomous vehicles, where LLMs could process sensory inputs, comprehend their surroundings, and consequently make more informed decisions. Building on these insights, additional studies \cite{bian_chatgpt_2023-1}, \cite{zhu_ghost_2023}, \cite{yao_react_2023}, \cite{shinn_reflexion_2023},  and \cite{huang_instruct2act_2023} have further enriched our understanding of LLMs' capabilities, underscoring their potential in decision-making, reasoning, and synergizing reasoning and acting.

\section{Perspective: the Role of LLMs in Advancing Autonomous Vehicles}
\label{sec:perspective}

As previously established in the earlier section, we've established that LLMs serve as the ``brain," facilitating driver interaction and decision-making, while the useful sensory tools and actuation function as the vehicle's ``eyes" and ``hands" respectively. To be more specific, When a driver requests a particular operation, the LLM prompts the related modules to provide data that has been processed to extract relevant information from the environment. By integrating the linguistic analysis of LLMs with the processed sensory inputs from the selected modules, the LLM can then make well-informed decisions. If the command is deemed both feasible and safe based on the prior analysis, the LLMs will transmit the corresponding instructions to the vehicle's controller. This includes components such as the steering wheel, throttle pedal, braking, and other control elements, enabling them to execute the necessary operations. Alternatively, if the operation is deemed inappropriate, the LLMs will provide drivers with a detailed explanation as to why the requested action is not suitable for execution.

Revisiting the initial example, when drivers issue the command to overtake the vehicle ahead, the LLMs come into play by querying the perception module for pertinent processed information. This includes details such as the distance and speed of the target vehicle, the velocity of the ego vehicle, road conditions of potential lanes, and the presence of other vehicles and their distances on those lanes. Through an analysis of the provided data and the given command, the LLMs decide whether to execute the driver's request. If the decision is affirmative, the LLMs subsequently communicate instructions to the controller, guiding the next course of action.

Having explored this intricate interaction between LLMs and the autonomous vehicle's decision-making process, we shift our focus to a broader context and propose the concept of a human-centric LLMs integrated framework for autonomous vehicles. As shown in \cref{fig:framework}, the physical world comprises human drivers, vehicles, and traffic objects. In the physical world, human drivers are the central agents in the physical world, sending commands and instructions to LLMs as they navigate roadways. The traffic environment contains various elements including vehicles, pedestrians, traffic lights, road conditions, and traffic cones, all of which contribute to the complexity of movement and interactions on the road. The vehicle, directed by the LLMs, operates within this ecosystem, executing the commands it receives from either drivers or LLMs through controllers and actuators.

The virtual world \cite{8944077} includes LLMs, memory, and essential tools which include the perception module, localization module, and in-cabin monitor. The perception module acquires raw input from sensors, including external cameras, LIDARs, and radars \cite{9398692,9506829,9882982,cui_radar_2023}, and processes this data into a format suitable for the LLMs. The localization module employs GNSS data to determine the vehicle's precise location. Within the vehicle, the in-cabin monitor employs internal cameras, thermometers, and other sensors to vigilantly observe the in-cabin environment, preempting distractions, extreme temperatures, or uncomfortable conditions. At the core of the entire framework lies the LLMs, serving as its central intelligence. They receive commands from drivers, subsequently initiating queries to pertinent modules for related information. Furthermore, the memory section acts as a repository, storing historical operations and drivers' preferences, enabling continuous learning and enhancement for the LLMs. This repository of experiences equips the LLMs to make analogous decisions when confronted with similar situations, bolstering the system's adaptability and performance over time. the memory also houses maps and local law information, leading the LLMs to make even wiser decisions adaptable to a variety of scenarios.

\begin{figure*}[t!]
  \centering
  \includegraphics[width=\textwidth]{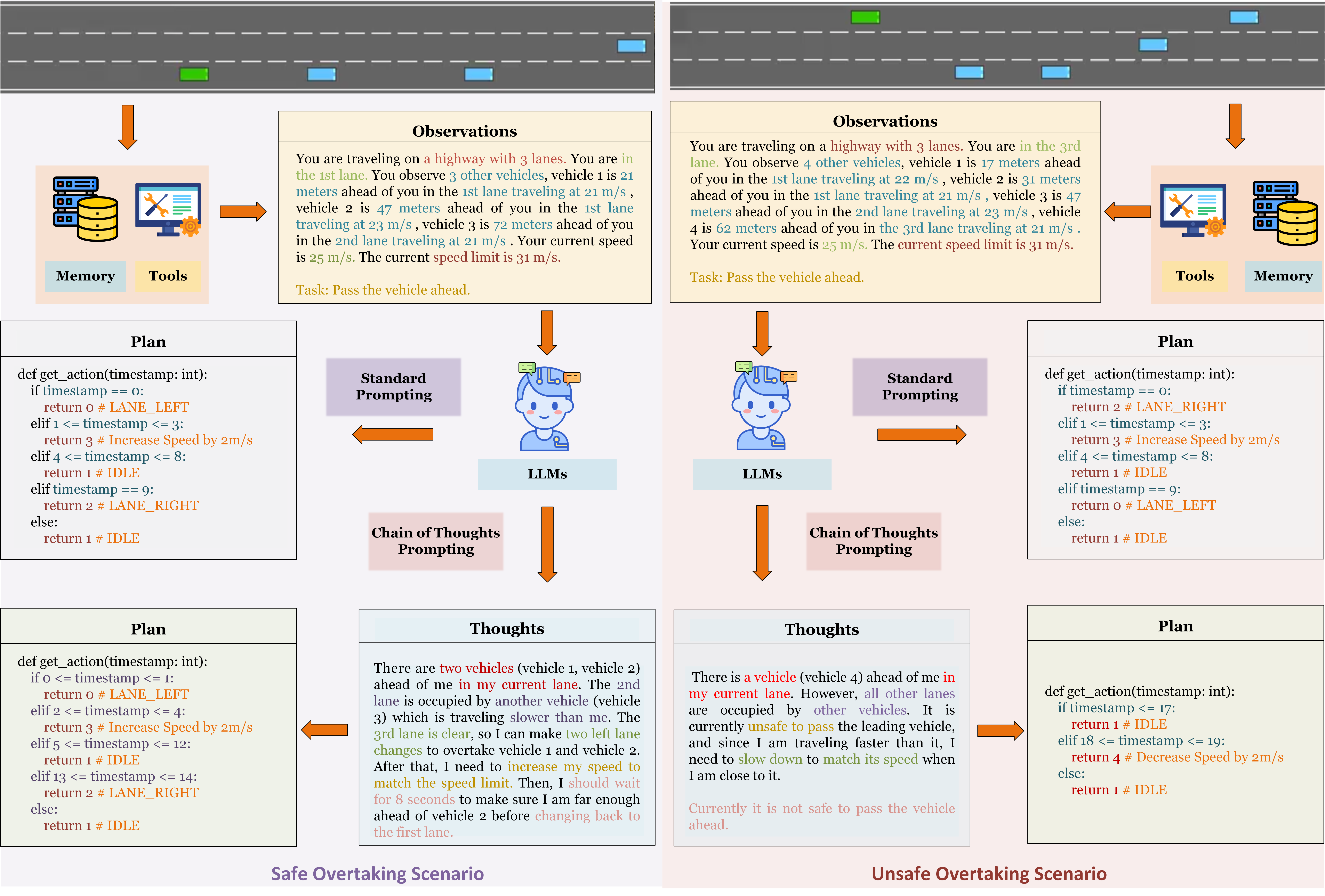}
  \caption{LLMs' decision-making workflows using both standard prompting and chain-of-thought prompting in the highway scenario}
  \label{fig:highway_scenario}
\end{figure*}

\section{Experiments: Cases Study with GPT-4}
\label{sec:experiments}

To gain a deeper understanding of the practical capabilities of LLMs in the context of autonomous driving tasks, we embark on an insightful exploration involving complex decision-making scenarios. This comprehensive case study serves as a compelling demonstration of how LLMs can effectively enhance autonomous driving cars to replicate decision-making processes. Our investigation is structured in two distinct phases. We design and present several complex real-world situations to assess the decision-making proficiency of LLMs. This section covers an in-depth understanding of these case studies, which serves to underscore the practical implications of leveraging LLMs for enhanced autonomous driving.


\subsection{Simulation Implementation Details}
For our study, we leverage GPT-4 to execute a closed-loop driving experiment in HighwayEnv, a collection of environments for autonomous driving and tactical decision-making tasks \cite{highway-env}. Our primary objective was to validate the LLMs' capabilities in areas of interpretation, chain-of-thought reasoning, personalization ability, and interaction with environmental conditions. The LLMs cannot directly access environmental data. To bridge this gap, we employed the tools and memory modules we mentioned before to extract environmental information from the simulation. Utilizing tools such as perception modules to capture environmental information and localization modules to determine the vehicle's position, alongside a memory storing local laws and traffic data, ensures LLMs with a well-rounded understanding of current situations. The output from them serves as the observations in the experiments and was then provided to LLMs for further processing, reasoning, and decision-making.

Our experiment for the chain-of-thought prompting was compared to two distinct experimental setups. The first was characterized by feeding examples that only include the environmental observations and their corresponding solutions. This was intended to measure the LLMs' performances in complex driving scenarios based solely on their foundational training. Conversely, the second setup is termed the ``chain-of-thoughts" group. In this approach, not only were the observations and solutions relayed to the LLMs, but they were also provided with a comprehensive breakdown of our decision-making process. This setup was designed to show the potential enhancements in performance, adaptability, and interpretation when LLMs are bolstered by in-context learning and chain-of-thought prompting.

Additionally, we also explore the potential for personalization in merging scenarios by establishing three groups with different queries: one with no queries provided, another instructed to ``drive more aggressively," and the last one instructed to ``drive more conservatively. Through this setup, we aim to analyze how varying queries or commands may influence LLMs' equipped vehicles' behavior and decisions during highway merging, with a particular focus on uncovering the potential for personalization in driving experiences and safety dynamics.

\subsection{Highway Overtaking}

In our simulation setup, the environment is designed to simulate real-world driving conditions for an autonomous vehicle. The vehicle is inherently equipped with Lane Keeping Assist (LKA), a feature that is perpetually activated, ensuring the vehicle maintains its position within the confines of its current lane. Lane identification is structured in a way that counting initiates from the rightmost lane, assigning it the lowest lane ID. The vehicle is afforded five distinct operational actions. These are 0, enabling a leftward lane change ('LANE LEFT'); 1, maintaining the current state or position ('IDLE'); 2, facilitating a rightward lane change ('LANE RIGHT'); 3, accelerating the current speed by 2 meters per second; and 4, reducing the speed by the same magnitude. This structured environment and action set are designed to test the LLM's proficiency in varied highway scenarios.

In our closed-loop driving experiment on HighwayEnv \cite{highway-env} using GPT-4, we try to assess the LLMs' interpretation, chain-of-thought, and environmental interaction abilities. Our experimental design includes two distinct highway scenarios. In the first one, the environment was safe for overtaking the vehicle ahead; in contrast, the second scenario presented conditions where overtaking was considered unsafe and not suitable. The emphasis was on observing the LLMs' reactions and decision-making in various conditions. For each scenario, we employ two distinct training methods. One method utilized standard prompting for training, and the other used the chain-of-thought prompting approach. Through this design, our objective was to discern and highlight the comparative advantages of using chain-of-thought prompting over standard prompting techniques. The whole working process for safe and unsafe scenarios is shown in Fig. \ref{fig:highway_scenario}  respectively. When prompted using the chain-of-thoughts method, the LLMs first generate comprehensive and reasoned thoughts before suggesting a driving plan. In contrast, with the standard prompting training method, the LLMs directly present the plan. The plans derived from these two methods have distinct differences.

\begin{figure}
\centering
\begin{subfigure}{\linewidth}
    \includegraphics[width=\textwidth]{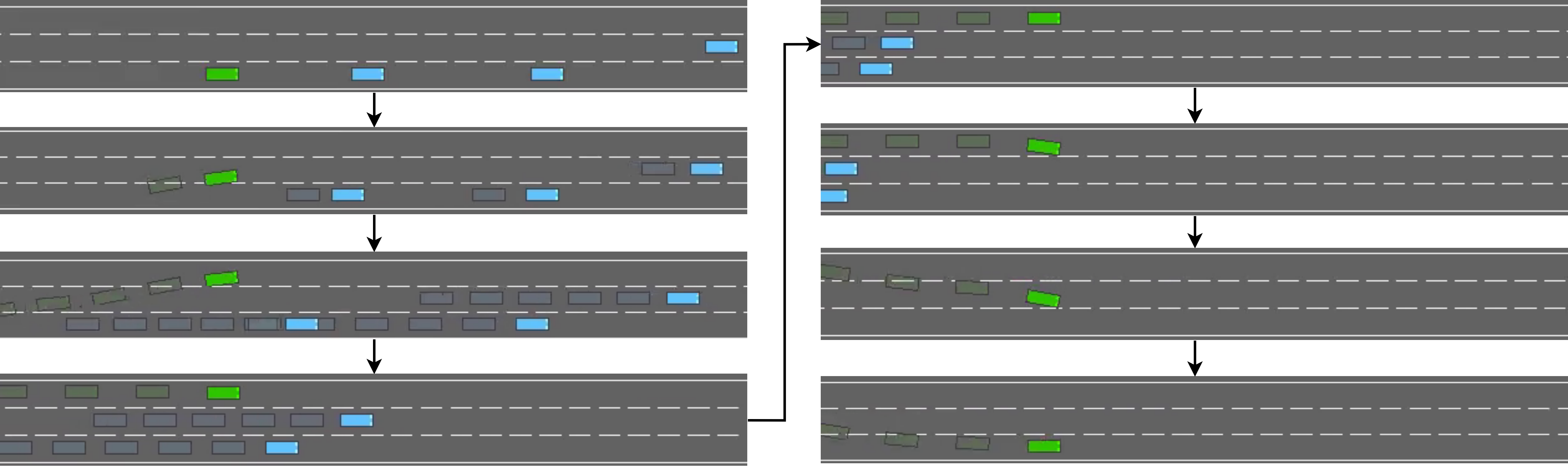}
    \caption{Training LLMs with the chain-of-thought reasoning method}
    \label{fig:high_safe_chain}
\end{subfigure}
\hfill
\begin{subfigure}{\linewidth}
    \includegraphics[width=\textwidth]{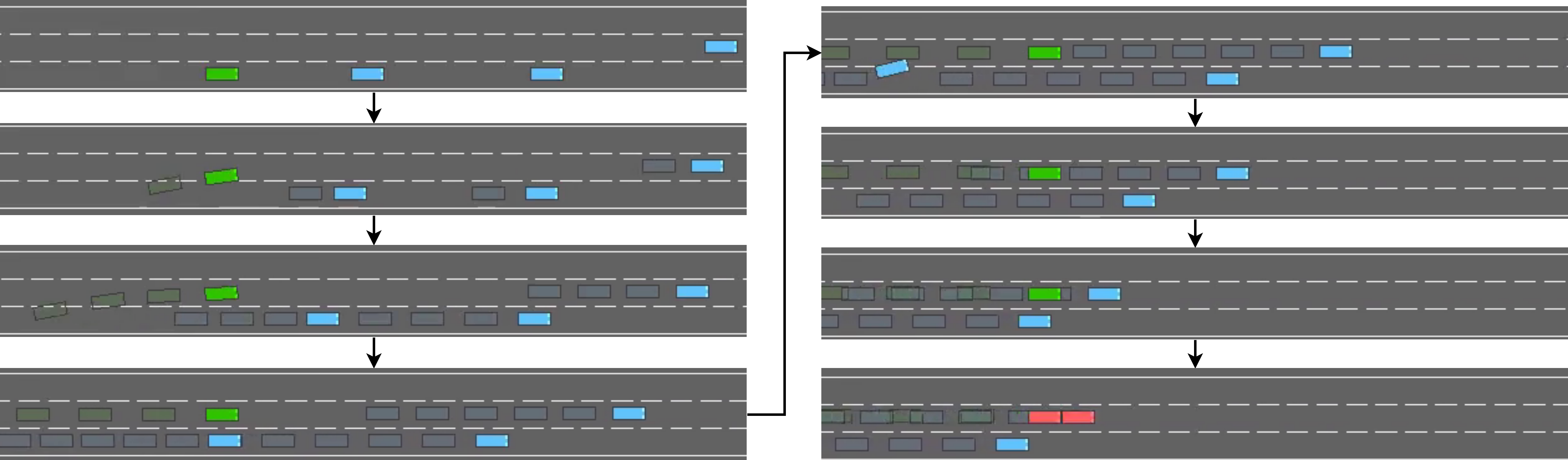}
    \caption{Training LLMS with standard prompting method}
    \label{fig:high_safe_standard}
\end{subfigure}
        
\caption{Visualization of experiment results for safe overtaking scenarios in the highway environment, where the green vehicle is the ego vehicle}
\label{fig:highway_safe_experiments}
\end{figure}

As previously mentioned, in our highway scenario testing, we evaluate two specific situations: one where it is regarded safe to overtake the vehicle in front, and another where it is considered unsafe to do so. The experiment results for the safe scenario are shown in Fig. \ref{fig:highway_safe_experiments} while those for the unsafe one are shown in Fig. \ref{fig:highway_unsafe_experiments}. Our experiments have revealed significant distinctions between LLMs prompted using the 
chain-of-thought prompting method and those trained using standard prompting. In the case of chain-of-thought prompting, the vehicle can intelligently assess the safety of overtaking the front vehicle. If the conditions are regarded safe, LLMs trained through this method will execute the most reasonable actions: they observe neighboring lanes for clearance, change lanes, accelerate to the speed limit to overtake, and subsequently return to their original lane when the lanes are clear. Conversely, if the situation is assessed as unsafe, these LLMs instruct the vehicle to accelerate to the speed limit. As the vehicle approaches the front one, it continuously evaluates the situation for overtaking opportunities. Upon determining that overtaking is not safe, it promptly reduces speed and abandons the overtaking attempt.

In contrast, LLMs trained via standard prompting show a marked inability to adapt and make informed and wise decisions. These models consistently make reckless choices in both safe and unsafe scenarios, often resulting in accidents. This stark contrast in behavior underscores the effectiveness and safety advantages of the chain-of-thought prompting training approach for autonomous vehicles.

\subsection{The Merging Senario}
Drivers may adjust their driving behavior, becoming either more aggressive or cautious, influenced by numerous factors~\cite{9210586,9511852}. In this experimental framework, imagine traveling on a route that is part of your daily commute, and upon the initial experience, you find the autonomous driving behavior too aggressive for your preference. This situation establishes the context for interactive modulation of driving styles through verbal instructions to the LLMs. For instance, by stating ``drive more conservatively" to the LLMs, you can trigger a change in the autonomous driving behavior towards a more conservative mode. This change is aimed at aligning the driving style closer to your comfort zone, which enhances the user experience.

On a different day, if you are running late and desire a swifter transit, instructing the LLMs with a ``drive more aggressively" command would prompt a shift to faster and more aggressive driving behavior. This scenario illuminates the potential for dynamic adjustments in driving styles based on real-time user feedback. Through continuous input from the driver, the LLMs are anticipated to gradually discern an optimal driving style, behavior, or a set of driving-related parameters that resonate with the driver's preferences. This continuous learning and adaptability highlight the potential for LLMs to substantially augment personalized driving experiences in autonomous vehicles.

In our experimentation within the HighEnv environment, similar to the previous highway scenario, we utilized the LLMs with three distinct queries: ``Drive conservatively," ``Drive aggressively," and a no-query condition where the LLMs assumed control over the autonomous vehicles without any explicit instructions, the results are shown in \ref{fig:merging_scenario}. Our primary objective was to ascertain the LLMs' proficiency in comprehending and reasoning the drivers' commands, and subsequently, reacting in alignment with these commands. The variation in driving strategies among these queries provided a fertile ground for evaluating how well the LLMs could personalize the driving experience based on the given instructions.

In the ``Drive conservatively" query, we aim to observe a cautious driving approach by the LLMs, anticipating behaviors such as maintaining lower speeds and avoiding risky overtaking maneuvers. Conversely, under the ``Drive aggressively" instruction, we seek to see a more aggressive driving style, with the LLMs potentially engaging at faster speeds and undertaking overtaking maneuvers where possible. The no-query condition serves as a control to illustrate the LLMs' default driving behavior without explicit driver instructions. Through these varying instructions, we intend to explore the extent to which LLMs can personalize driving experiences based on drivers' preferences. This experimental design stands as a significant effort towards understanding the potential of LLMs in fostering a more intuitive and user-centric autonomous driving solution, by aligning vehicle behavior closely with drivers' commands.

\begin{figure}[t!]
\centering
\begin{subfigure}{\linewidth}
    \includegraphics[width=\textwidth]{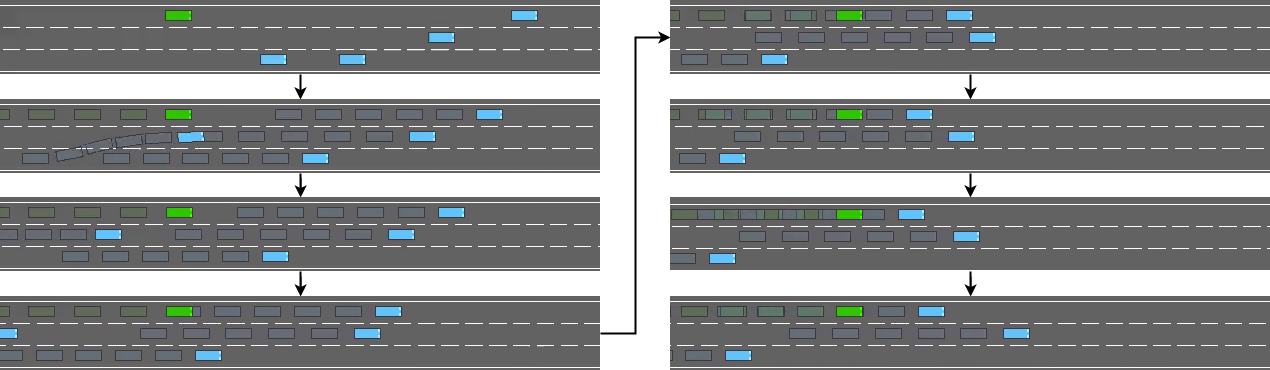}
    \caption{Training LLMs with chain-of-thought reasoning method}
    \label{fig:high_safe_chain}
\end{subfigure}
\hfill
\begin{subfigure}{\linewidth}
    \includegraphics[width=\textwidth]{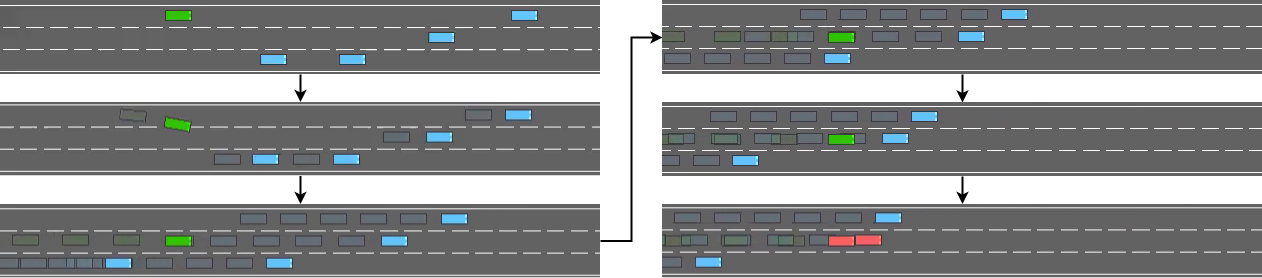}
    \caption{Training LLMs with standard prompting method}
    \label{fig:high_safe_standard}
\end{subfigure}

\caption{Visualization of experiment results for unsafe overtaking scenarios in the highway environment, where the green vehicle is the ego vehicle}
\label{fig:highway_unsafe_experiments}
\end{figure}

\begin{figure}
\centering
\begin{subfigure}{\linewidth}
    \includegraphics[width=\textwidth]{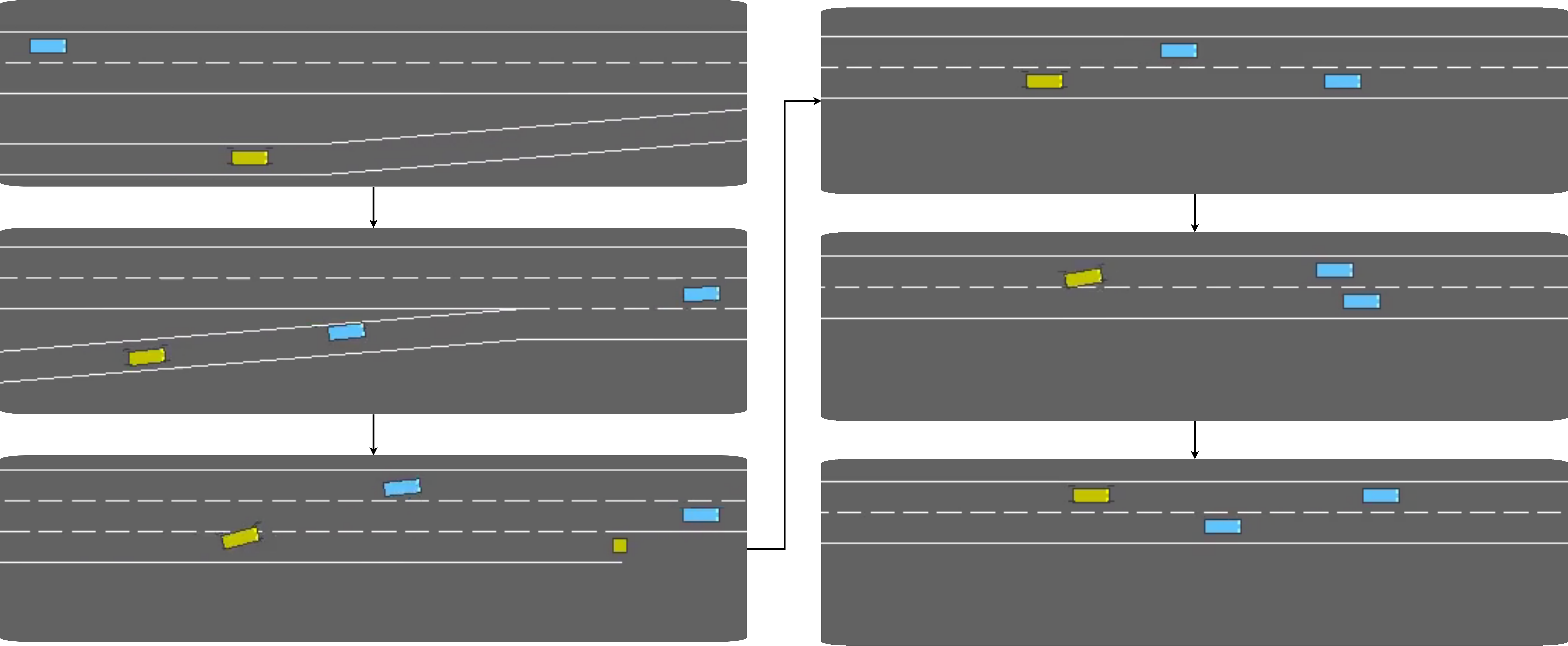}
    \caption{The merging scenario without command query}
    \label{fig:personal_regu}
\end{subfigure}
\hfill
\begin{subfigure}{\linewidth}
    \includegraphics[width=\textwidth]{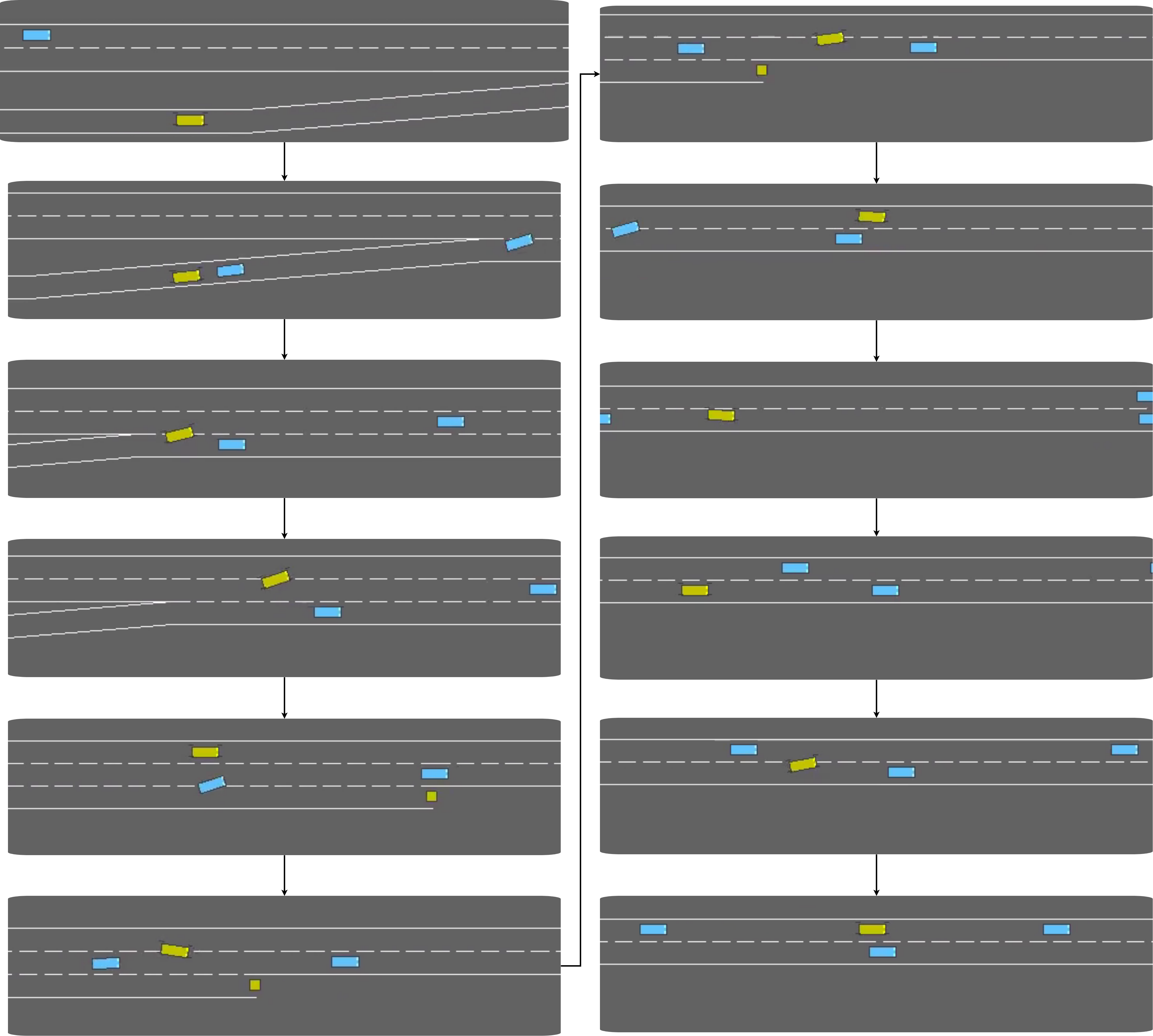}
    \caption{The merging scenario with 'drive aggressively' command}
    \label{fig:personal_agg}
\end{subfigure}
\hfill
\begin{subfigure}{\linewidth}
    \includegraphics[width=\textwidth]{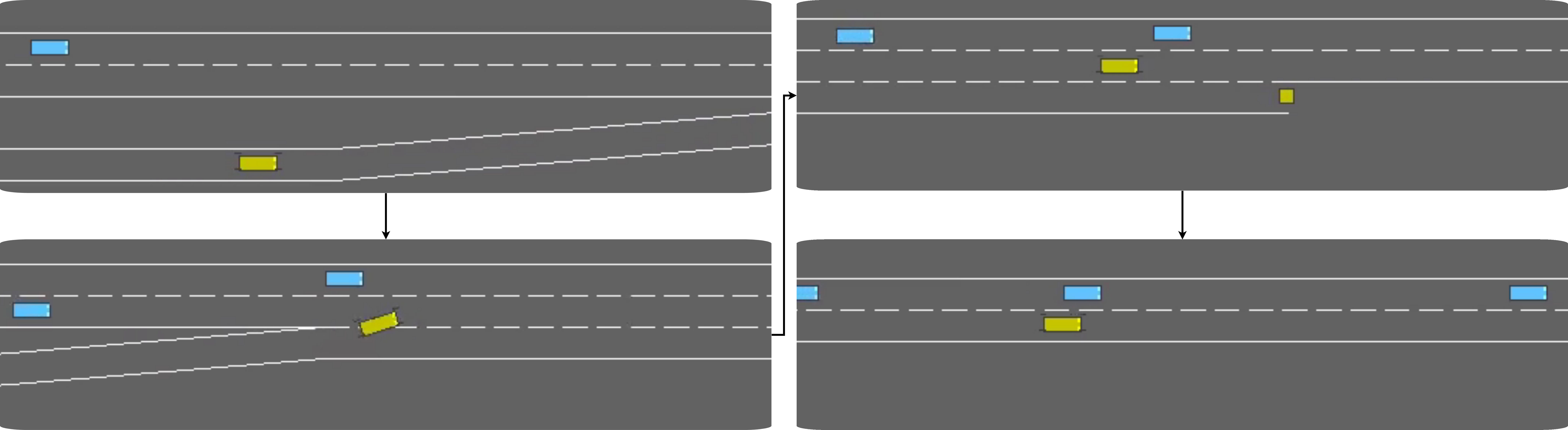}
    \caption{Merging scenario with 'drive conservatively' command}
    \label{fig:personal_conser}
\end{subfigure}
\caption{Visualization of experiment results for merging scenarios in the highway environment, where the yellow vehicle is the ego vehicle}
\label{fig:merging_scenario}
\end{figure}

As it is shown in Table \ref{tab:personal}, the results from the experiments illustrate a distinct difference in driving behaviors as commanded by the verbal instructions to the LLMs. When instructed to ``drive more conservatively," the LLMs exhibited a cautious driving style with notably lower acceleration, less aggressive steering inputs, lower top speed, greater front gap, and an overall longer time to complete the designated route, reflecting a more safety-oriented driving approach.

On the other hand, with the ``drive more aggressively" instruction, a more assertive driving behavior was manifested. This was marked by higher acceleration, more pronounced steering inputs, increased top speed, a smaller front gap, and a significantly reduced overall time to navigate the route, showcasing a more time-efficient but risk-prone driving stance.

In the scenario where no explicit command was given to the LLMs, a balanced driving behavior was observed, falling between the conservative and aggressive driving commands. Through these diverse scenarios, the ability of LLMs to dynamically adjust driving behaviors based on real-time user feedback is well-demonstrated, indicating a promising avenue toward achieving personalized driving experiences in autonomous vehicles.

\section{Analysis and Discussion}
\label{sec:discussion}
\subsection{In-Context Learning}
In our experiments involving highway overpassing and merging, we witnessed the power of in-context learning. When we provided the LLMs with specific examples and operational thoughts for these scenarios, they successfully navigated the challenges, underlining the importance of context. This shows that while LLMs have vast generalized knowledge, they can become powerful in autonomous driving when they're aligned with specific contexts. Our experiments also showcased the potential of in-context learning to enhance the predictive abilities of LLMs. In autonomous driving, accurate predictions are vital. By providing scenario-specific contexts, we used prompts to improve the LLM's predictive performance, leading it toward desired outcomes.

One of the most evident benefits of in-context learning in LLMs is its inherent adaptability. Traditional decision-making models often need retraining from scratch or employing a pre-trained model when faced with a new scenario.  Our highway overpassing and merging driving experiments exhibited the strengths of in-context learning. When the LLMs were equipped with more context details, they adeptly navigated these intricate scenarios, avoiding the crash scenarios observed when no context was provided. Traditional decision-making models often need retraining from scratch or employing a pre-trained model when faced with a new scenario, but the in-context learning in our LLMs simply required relevant guidance to recalibrate their decision-making approach. Also, the financial cost of constantly retraining traditional models can be expensive, especially considering the computational resources required. In-context learning in LLMs presents a more cost-effective alternative. By simply providing contextual guidance, we can recalibrate the model's behavior, reducing both computational and financial overheads.

\subsection{Reasoning}

The LLMs were tasked with processing this multilayered data sourced from the perception module (vehicle speeds and distances), the localization module (road and environmental conditions), and the in-cabin monitoring system (driver's attention level and safety measures like seatbelts). The LLMs formulated a comprehensive motion plan that prioritized safety while efficiently executing the driver's command to overtake the front vehicle.

In the experimental scenarios, the LLMs showed their advanced reasoning ability by not just collecting and analyzing data but also applying layers of context-sensitive reasoning. The LLM evaluated the speeds and distances of surrounding vehicles and even the traffic conditions to determine the safest and most efficient trajectory for overtaking. This capability to reason in real-time, considering multiple factors dynamically, significantly contributes to road safety and operational efficacy. The LLM didn't merely follow pre-defined rules but adapted its decision-making to the unique circumstances, highlighting its potential for enhancing the future of autonomous driving.

\begin{table*}[!ht]
\centering

\caption{Comparison of Driving Behaviors on Varied Verbal Instructions (\# Means Number of)}
\begin{tabular}{@{}l|cccccc@{}}
    \toprule
    \multicolumn{1}{c}{Commands} \vrule & Mean Abs Acceleration & Mean Abs Steering  & Max Abs Speed  & Min Front Gap & Overall Time & \# Chaging Lanes\\
    \midrule
    \multirow{1}{*}{Drive more Aggresively} 
    & 3.10 $m/s^2$ & 0.03 rad & 34.77 $m/s$ &7.17 m &24.33 s &6\\
    \midrule
    \multirow{1}{*}{Drive more Conservatively} 
    &  0.18 $m/s^2$ & 0.01 rad & 20.00 $m/s$ & 39.14 m & 46.20 s & 1\\
    \midrule
    \multirow{1}{*}{No Extra Command} 
    & 1.41 $m/s^2$ &0.02 rad &  27.43 $m/s$ & 24.01 m & 34.20 s & 2\\

    \bottomrule
\end{tabular}
\label{tab:personal}
\end{table*}

\subsection{Chain-of-Thoughts}
The experiments conducted in the scenarios of highway overpassing and driving in merging situations illuminate a critical facet of LLMs when applied to autonomous driving: their dependency on contextual input for optimized performance. Without an initial framework or examples to guide decision-making processes, LLMs are prone to taking suboptimal actions, which caused vehicle crashes in our experiments.

This phenomenon can be attributed to the underlying architecture and training methodology of LLMs. Traditionally, these models excel in tasks where they've been exposed to vast amounts of relevant data during training. In situations where they face novel challenges or contexts without direct or related training, they can make decisions that seem weird or risky to human observers.

Introducing a chain-of-thought prompting - a structured sequence of reasoning or examples - to LLMs appears to bridge this knowledge gap. By presenting a series of logical and interconnected steps or directives, the model can better navigate the complexities of real-world driving scenarios.

In our experiments, the chain-of-thought prompting provided to LLMs acted as a guiding signal, ensuring that the model was aligned with human-like reasoning and practical considerations of driving. The process can be likened to a human driver receiving step-by-step instructions or guidance when learning a new driving maneuver. Over time, with consistent exposure to these chains of thoughts, LLMs could potentially generalize these guidelines across similar driving scenarios.

The observable enhancement in LLM performance, when equipped with chains of thoughts, underscores their potential in autonomous driving:
\paragraph{Robust Decision-making Framework:} The introduction of the chain-of-thought prompting in our experiments has revealed a distinct enhancement in the decision-making capabilities of LLMs. With a clear and contextual understanding provided through the chain-of-thought prompting, LLMs demonstrated more reliable and feasible performance, especially in intricate scenarios like highway overpassing and merging. This suggests that supplementing LLMs with a structure

\paragraph{Safety and Compliance with Legal Norms:} Incorporating safety constraints and legal guidelines into the LLMs' decision-making process not only prevents hazardous behaviors but also ensures compliance with road laws. As autonomous vehicles become a common sight on our roads, they must adhere strictly to both safety protocols and legal standards. Our approach using the chain-of-thought prompting with these constraints ensures that autonomous vehicles can navigate complex environments without endangering passengers or violating regulations.

\paragraph{Personalized Driving Experience:}The adaptability of LLMs with the chain-of-thought prompting framework offers a unique opportunity for personalization in autonomous driving. Each driver, or passenger in the context of autonomous vehicles, has a distinct preference when it comes to driving style, comfort, and responsiveness. By capturing these nuances through chain-of-thought prompting, LLMs can adjust driving behaviors to cater to individual preferences. Whether it's a smoother acceleration profile, particular lane preferences on highways, or preferred routes in cityscapes, the potential to customize the driving experience leads to a more comfortable and enjoyable journey for passengers.

\subsection{Personalization}
The merging scenario experiments show the potential of LLMs in personalizing the driving experience based on verbal instructions. The varying driving statistics based on verbal instructions - ``drive more conservatively", ``drive more aggressively", and no instruction - indicate a robust understanding and translation of these commands into user-centric actionable driving strategies by the LLMs. This personalized approach has the potential to enhance user satisfaction and safety in autonomous vehicle operations.

The distinct change in driving behavior in response to verbal instructions such as ``drive more conservatively" and ``drive more aggressively" shows a huge potential in understanding and execution of commands by the LLMs. Such ability to adjust driving behavior continuously from conservative to aggressive based on real-time feedback demonstrates the potential for a highly personalized driving experience. Over time, with continuous input from drivers, LLMs can potentially fine-tune the vehicle's driving parameters to align closely with individual driver preferences, thus providing a tailored driving experience.

The potential for real-time adaptation to verbal instructions brings a significant advantage in improving autonomous vehicle systems to become more user-centric. Such a level of personalization can significantly improve user satisfaction by allowing individuals to determine the aggressiveness or conservativeness of the driving style based on their current preferences or external conditions.

Furthermore, this real-time feedback loop between the user and the LLMs could be extended to learn and anticipate individual preferences over time, creating a truly personalized driving policy. This could lead to the development of a profile for each user, allowing the LLMs to know the driver's preferences even before they provide instructions. Such a level of personalization could extend beyond driving behavior to include other aspects like control of the in-vehicle environment, route preferences, and more, thus making the autonomous driving experience more intuitive and enjoyable.

\subsection{Interpretation}
The language interaction capabilities of the LLM proved crucial for trust-building. When the driver commanded to ``pass the vehicle in front," the LLM assessed various factors and clearly communicated its reasoning to the driver. This transparent interaction not only enhanced safety but also instilled greater confidence in the vehicle's autonomous capabilities. This crucial advantage enhanced transparency and trust. When the vehicle makes a complex decision, such as overtaking another vehicle on a high-speed, two-lane highway, passengers and drivers might naturally have questions or concerns. In these instances, the LLM doesn't just execute the task but also articulates the reasoning behind each step of the decision-making process. By providing real-time, detailed explanations in understandable language, the LLMs reveal the vehicle's actions and underlying logic. This not only satisfies the innate human curiosity about how autonomous systems work but also builds a higher level of trust between the vehicle and its occupants.

\section{Conclusions and Future Work}
\label{sec:conclusion}

\subsection{Future Work}
 The integration of LLMs into autonomous vehicles is an exciting and promising area, but it also leaves several questions open for future research. One of the most important directions is the fine-tuning of these LLMs for specific tasks within the vehicle's operational framework. For example, LLMs could be tailored to better understand and interpret perception data, providing more accurate and timely decision-making inputs for motion planning tasks

 A vital field for future work is the establishment of a benchmark to evaluate the capabilities of advanced LLMs in autonomous driving tasks. Currently, the absence of such a standardized evaluation system prevents objective comparisons between models. This benchmark would focus on key tasks for LLMs within autonomous vehicles—ranging from voice command recognition to data-driven motion planning—and create both quantitative and qualitative metrics for assessment. This benchmark would serve as a foundational tool for researchers and an evolving platform that reflects the state-of-the-art in the field, facilitating the accelerated development and deployment of LLMs in autonomous driving systems.
 
 Another critical area is the improvement of real-time decision-making. While our framework shows the capability of LLMs to process vast amounts of data and provide intelligent feedback in a short time, more research is needed to make these processes more efficient and robust in real-world applications.

Finally, the ethical and social implications of LLM-driven autonomous vehicles also need to be explored. As these models are given more responsibility for decision-making, issues around accountability and data privacy will need to be considered. Future work could aim to establish frameworks for ethical considerations specific to the application of LLMs in the automotive space.

\subsection{Conclusion}

In conclusion, our paper has provided a comprehensive framework for integrating LLMs into the ecosystem of autonomous vehicles. We have highlighted how LLMs offer unparalleled reasoning capabilities that can make autonomous systems more flexible and responsive to complex, real-world scenarios. This is a significant advantage over traditional autonomous systems, which often rely on static decision-making algorithms. Additionally, by leveraging the capabilities of LLMs, we can enrich the human-car interaction, providing a more reliable, intuitive, and responsive interface. Unlike traditional autonomous systems, which often lack the capacity for natural language understanding, LLMs can handle complex requests, offer real-time feedback, and comprehensive explanations, and even assist in decision-making during complex or rare driving scenarios.

Our experiments verified the utility of integrating LLMs in various modules of an autonomous vehicle, such as decision-making, motion planning, and human-vehicle communication. The results indicate a promising pathway towards the utilization of LLMs in enhancing not only the safety but also the user experience in autonomous vehicles. However, some future directions such as fine-tuning LLMs for specific vehicular tasks, developing a comprehensive benchmark for evaluating LLMs, improving real-time decision-making efficiency, and addressing the ethical and social implications of using LLMs are still need to be further explored.

Overall, the findings point to a future where LLMs could play a significant role in the progression of autonomous vehicle technology, offering a combination of efficiency, safety, and user-centric design that traditional systems have yet to achieve.

\ifCLASSOPTIONcaptionsoff
  \newpage
\fi



\bibliographystyle{IEEEtran}
\bibliography{bib/LLM_ref,bib/ma,bib/new}

\begin{IEEEbiography}
[{\includegraphics[width=1in,height=1.25in,clip,keepaspectratio]{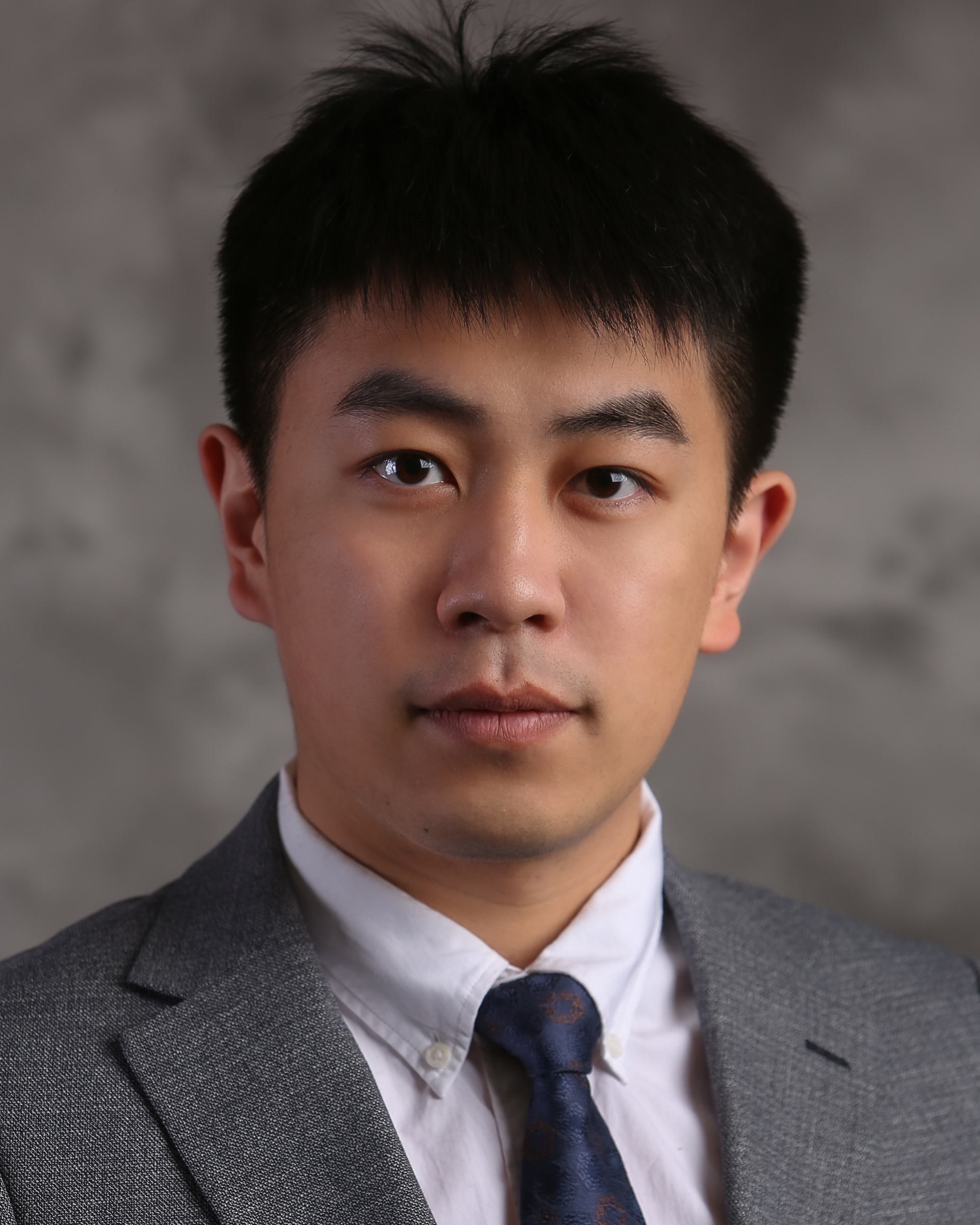}}]
{Can Cui}
(S'23) received his M.S. degree in Electrical and Computer Engineering from the University of Michigan, Ann Arbor in 2022. Currently, he is a Ph.D. student in the College of Engineering at Purdue University. His research pursuits encompass autonomous driving, Advanced Driver-Assistance Systems (ADAS), applied machine learning, controls, and digital twin.
\end{IEEEbiography}
\vfill

\begin{IEEEbiography}[{\includegraphics[width=1in,height=1.25in,clip,keepaspectratio]{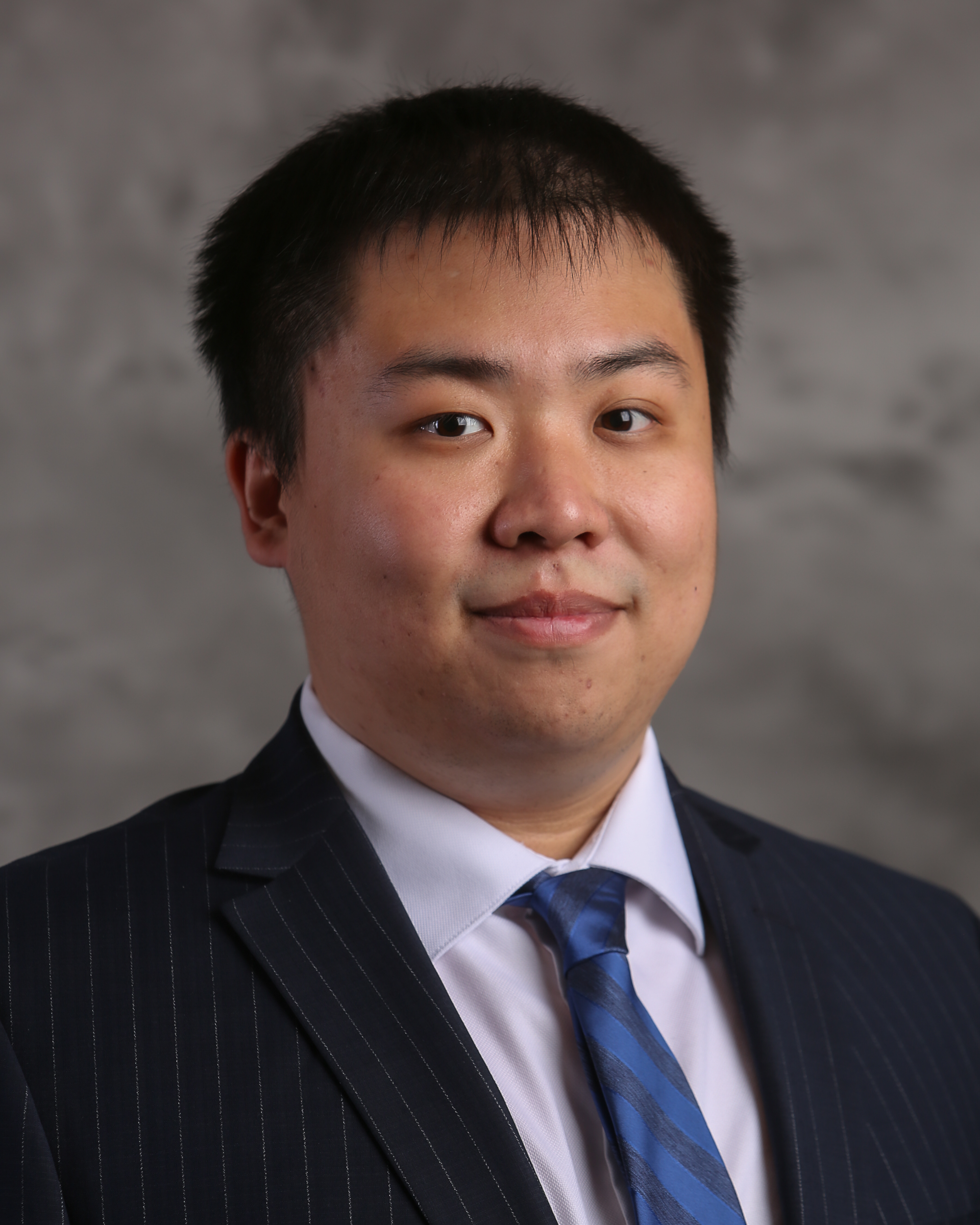}}]{Yunsheng Ma}
(S'23) earned his M.S. degree in Computer Science from New York University in 2022. Currently, he is pursuing a Ph.D. degree in the College of Engineering at Purdue University, specializing in applied AI/ML, autonomous driving, and digital twin. His research has been published in esteemed conferences such as AAAI, UAI, CVPR, and ITSC.
\end{IEEEbiography}

\vfill
\begin{IEEEbiography}[{\includegraphics[width=1in,height=1.25in,clip,keepaspectratio]{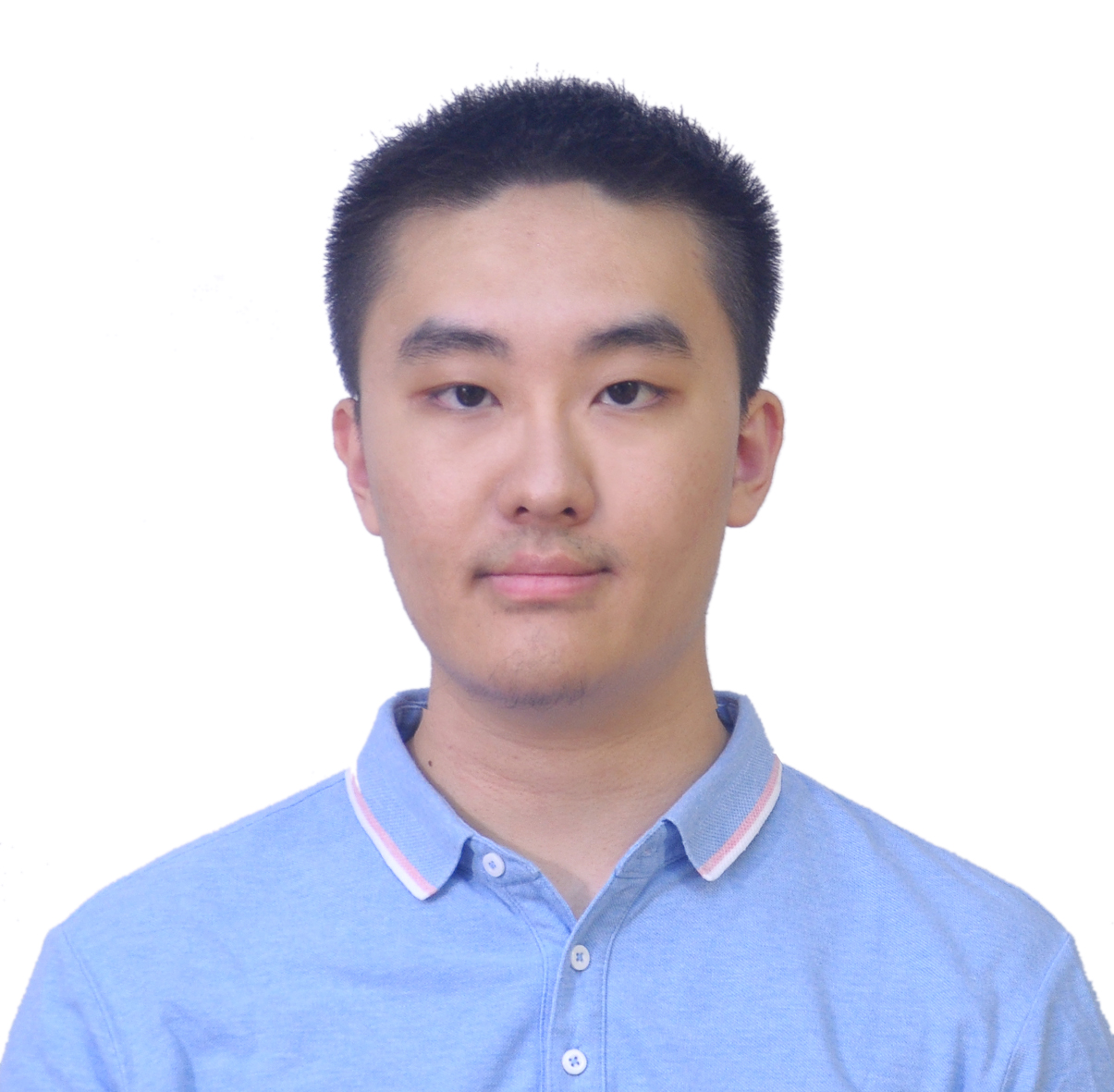}}]{Xu Cao}
(S'18-M'23) received his M.S. in Computer Science from New York University in 2022 and his B.S. from Fudan University in 2020. He is the co-founder of PediaMed.AI Lab. His research interests include AI for healthcare, AI for social good, autonomous driving, computer vision. His research achievements have been published in multiple top-tier conferences, including AAAI, IJCAI, ICASSP, UAI, BIBM with one of his co-authored papers earning the IAAI Application Innovation Award.
\end{IEEEbiography}
\begin{IEEEbiography}[{\includegraphics[width=1in,height=1.25in,clip,keepaspectratio]{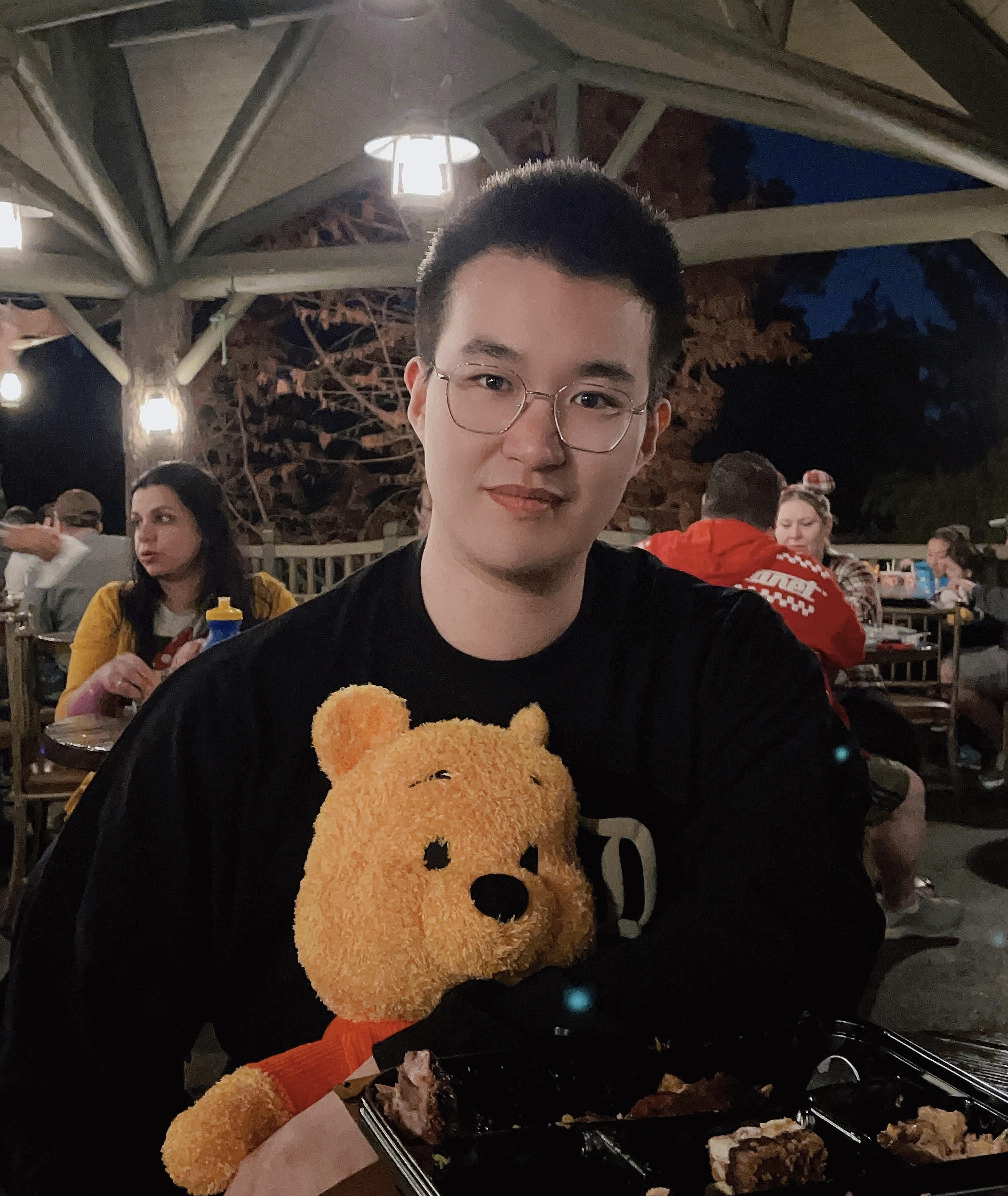}}]{Wenqian Ye}
(S'18-M'23) received his M.S. degree in Computer Science from New York University in 2022 and his B.S. degree from University of Illinois at Urbana Champaign in 2020. He is now pursuing a Ph.D. degree in the Department of Computer Science at University of Virginia, with a focus on Bayesian Machine Learning, Robustness, AI for Healthcare and Robotics. He has publications in top-tier conferences including UAI, ICASSP, and ITSC.
\end{IEEEbiography}

\begin{IEEEbiography}
[{\includegraphics[width=1in,height=1.25in,clip,keepaspectratio]{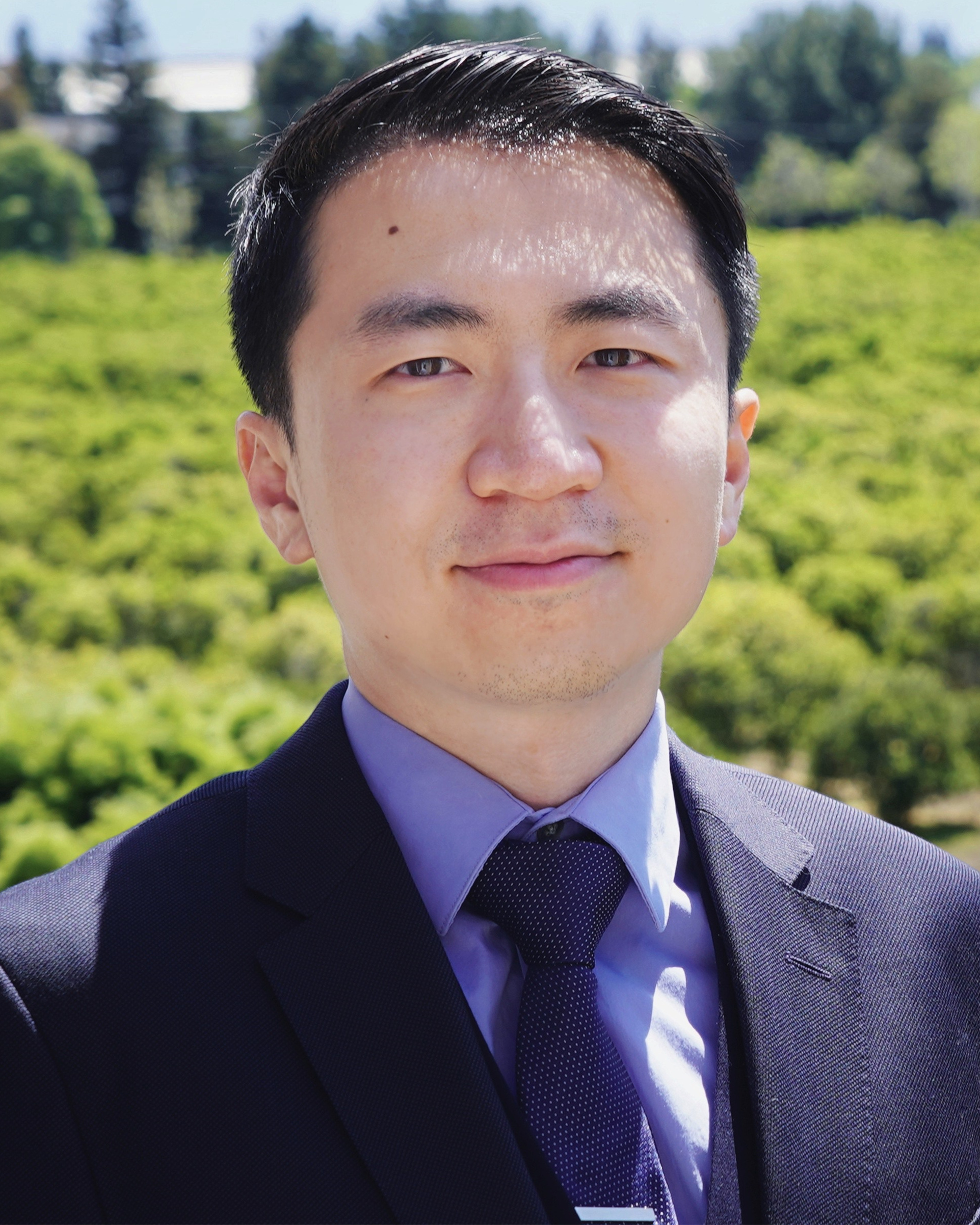}}]
{Ziran Wang}
(S'16-M'19) received his Ph.D. degree in Mechanical Engineering from the University of California, Riverside in 2019. He is a tenure-track assistant professor in the College of Engineering at Purdue University, where he leads the Purdue Digital Twin Lab. Prior to this, Dr. Wang was a principal researcher at Toyota Motor North America in Mountain View, California. 

Dr. Wang serves as the founding chair of IEEE Technical Committee on Internet of Things in Intelligent Transportation Systems, a member of three other IEEE technical committees, and a technical program committee member of multiple IEEE and ACM conferences. Dr. Wang is an associate of four academic journals, including IEEE Transactions on Intelligent Vehicles and IEEE Internet of Things Journal. His research achievements have been demonstrated at Consumer Electronics Show (CES), and acknowledged by the U.S. Department of Transportation Dissertation Award, the IEEE ``Shape the Future of ITS'' 1st Prize Award, and five other best paper awards from IEEE and SAE. He is an author of four book chapters, 50+ refereed papers, and 50+ patent applications. His research focuses on autonomous driving, human-autonomy teaming, and digital twin.
\end{IEEEbiography}

\end{document}